\documentclass[fleqn,usenatbib]{mnras}

\usepackage{newtxtext,newtxmath}

\usepackage[T1]{fontenc}

\DeclareRobustCommand{\VAN}[3]{#2}
\let\VANthebibliography\thebibliography
\def\thebibliography{\DeclareRobustCommand{\VAN}[3]{##3}\VANthebibliography}

\usepackage{graphicx}	
\usepackage{amsmath}	

\let\oldAA\AA
\renewcommand{\AA}{\text{\normalfont\oldAA}}

\title[UVIT view of NGC 628]{Understanding the secular evolution of NGC 628 using UVIT}

\author[K. Ujjwal et al.]{
K. Ujjwal,$^{1}$\thanks{E-mail: ujjwal.krishnan@res.christuniversity.in}
Sreeja S. Kartha,$^{1}$
Smitha Subramanian,$^{2}$
Koshy George,$^{3}$
Robin Thomas,$^{1}$
\newauthor
and
Blesson Mathew$^{1}$
\\
$^{1}$Department of Physics and Electronics, CHRIST (Deemed to be University), Bangalore 560029, India\\
$^{2}$Indian Institute of Astrophysics, Sarjapur Road, Koramangala, Bangalore 560034, India\\
$^{3}$ Faculty of Physics, Ludwig-Maximilians-Universität, Scheinerstr 1, Munich, 81679, Germany \\
}

\date{Accepted XXX. Received YYY; in original form ZZZ}

\pubyear{2022}

\begin{document}
\label{firstpage}
\pagerange{\pageref{firstpage}--\pageref{lastpage}}

\maketitle

\begin{abstract}
Secular and environmental effects play a significant role in regulating the star formation rate and hence the evolution of the galaxies. Since UV flux is a direct tracer of the star formation in galaxies, the UltraViolet Imaging Telescope (UVIT) onboard ASTROSAT enables us to characterize the star forming regions in a galaxy with its remarkable spatial resolution. In this study, we focus on the secular evolution of NGC 628, a spiral galaxy in the local universe. We exploit the resolution of UVIT to resolve up to $\sim$ 63 pc in NGC 628 for identification and characterization of the star forming regions. We identify 300 star forming regions in the UVIT FUV image of NGC 628 using ProFound and the identified regions are characterized using Starburst99 models. The age and mass distribution of the star forming regions across the galaxy supports the inside-out growth of the disk. We find that there is no significant difference in the star formation properties between the two arms of NGC 628. We also quantify the azimuthal offset of the star forming regions of different ages. Since we do not find an age gradient, we suggest that the spiral density waves might not be the possible formation scenario of the spiral arms of NGC 628. The headlight cloud present in the disk of the galaxy is found to be having the highest star formation rate density ($0.23 M_{\odot} yr^{-1} kpc^{-2}$)  compared to other star forming regions on spiral arms and the rest of the galaxy.
 
\end{abstract}




\section{Introduction}           
\label{sect:intro}

Star formation in a galaxy is considered one among the best aids to understand the evolution of the galaxy. In the nearby Universe, the evolution of galaxies is dominated by secular processes, whereas the violent processes are less common \citep{Kormendy2004}. Secular evolution is a process that results from the slow rearrangement of energy and mass. The presence of bars, oval disks, and spiral structures, interactions occurring inside the galaxy contributes to secular evolution \citep{Kormendy2013}. Instabilities occurring as a result of the presence of spiral arms play an important role in the secular evolution of a galaxy. Heating and radial migration of stars are the two secular processes attributed to the presence of spiral arms \citep{Lynden-Bell1972, Sellwood2011, Bautista2021}. Also, recent studies suggest that the arms could introduce gas inflow, and add to the secular evolution of the galaxy \citep{Kim2014, Baba2016}. It is worth noting that spiral galaxies make up most of the high star-forming galaxies in the blue cloud region of an optical color-magnitude diagram \citep{Baldry2004}. Since the star formation properties of the galaxies can be affected by the spiral structure, studying them will elucidate the understanding of the evolution of galaxies. 

The formation of spiral structures in the disk galaxies remains a question yet to be answered. Spiral density wave theory and swing amplification are the most accepted scenarios regarding spiral arm formation. According to the spiral density wave theory, static density waves constitute the long-lived stationary spiral arms \citep{Lin&shu1964}. On the other hand, swing amplification theory suggests that the local amplification in a differentially rotating disk results in the formation of spiral structure in a galaxy \citep{Goldreich1965, Julian1966, Elmegreen2011, Donghia2013}. Swing amplification theory considers spiral arms to be transient features.

Recent studies also explore the possibilities of whether the tidal interactions play a role in inducing the spiral features by generating localised disturbances which get enhanced by the swing amplification \citep{Kormendy1979, Bottema2003, Pettitt2016}. Along with that, the scenarios such as bar-induced spiral structure \citep{Contopoulos1980} and spiral features explained by a manifold \citep{Contopoulos1980, Athanassoula1992} are also considered as formation scenarios of spiral structures in galaxies. The longevity of spiral structure is the observational parameter that enables us to understand the formation scenario of spiral arms. 

In the spiral density wave consideration, at the corotation radius, the angular speed of stars and gas is equal to the angular pattern speed of the spiral features, whereas the material rotates faster than the pattern inside the corotation radius. Outside the corotation radius, material rotates slower compared to the pattern speed of the spiral structures. The gas, while entering into high dense regions of spiral arms, might experience a shock which could result in the star formation \citep{Roberts1969}. As they age, the stars move away from the spiral arms, resulting in an age gradient across the spiral arms. If we assume a constant angular speed for the spiral arms, comparatively younger star clusters will be located near the arm whereas the older star clusters will be found further away from the spiral arms. Hence the distribution of single stellar population equivalent age helps us to make a better understanding about the possible formation scenario for the spiral arms.

NGC 628 is an Sc-type galaxy at a redshift of z $\approx$ 0.00219 having an estimated distance of 9.6 Mpc \citep{Kreckel2018} with an inclination of $9^{\circ}$. NGC 628 is the most prominent member of a small group of galaxies. The group is centered on NGC 628 and the peculiar spiral NGC 660. Two well-defined spiral arms observed in optical and UV images make NGC 628 a typical example of grand design spirals. The galaxy properties are listed in Table \ref{tab:Table1}. NGC 628 has not gone through any recent interactions in the past 1 Gyr  \citep{Kamphuis1992}. This makes NGC 628 a good candidate to study the effects of secular evolution. The richness of ancillary data (Spitzer Nearby Galaxies Survey (SINGS); \citet{Kennicutt2003}, the GALEX Nearby Galaxy Survey (NGS); \citet{Gildepaz2007}, Legacy Extragalactic UV Survey (LEGUS); \citet{Calzetti2015}, Physics at High Angular Resolution in Nearby Galaxies (PHANGS); \citet{Leroy2021}, etc.), makes NGC 628 a perfect sample in a wide range of research topics including  galaxy evolution, stellar populations \citep[e.g.,][] {Cornett1994, shabani2018}, and star formation \citep[e.g.,][]{Elmegreen1983, Gusev2014}.

In this study, we intend to identify and characterize the star-forming regions' properties in NGC 628 using the FUV and NUV broadband observations of NGC 628 carried out using UVIT. Our primary goal is to understand the recent star formation activity in the grand design spiral galaxy NGC 628, to make a better understanding about the star formation in the spiral arms and hence to connect it with the possible spiral arm formation scenario. \citet{Gusev2013} exclusively studied the photometric properties of the spiral arms and the star clusters within NGC 628 using GALEX images. They classified the arms of NGC 628 into longer and shorter arms based on the extent of the arms. They considered the shorter arm as a distorted one. One of the significant conclusions from their study was that the longer arm of NGC 628 hosts a regular chain of star forming complexes, whereas the shorter arm did not. In a follow-up study, \citet{Gusev2014} suggested that the longer arm hosts slightly younger star forming regions than in the shorter arm.  \citet{shabani2018} studied the properties of spiral arms in NGC 628 using the stellar cluster catalog from the Legacy Extragalactic UV Survey (LEGUS) program and they suggested that there does not exist any age gradient across the arms of NGC 628. The advantage of our study is the better resolution and larger field of view provided by the UVIT images. \citet{Gusev2013} used the GALEX images, but the resolution of UVIT opens up an opportunity to revisit the UV properties of the galaxy with better resolution. Due to the field of view limitations, \citet{shabani2018} could not consider the spiral arms completely in their study. \citet{Jyoti2021} studied the star forming complexes in three nearby galaxies including NGC 628, using UVIT images. The study focused mainly on the properties of the star forming complexes inside and outside the optical radius ($R_{25}$). In this present study, the importance is given to the star formation properties of the spiral arms of NGC 628 to understand the formation mechanism behind the spiral arm formation.

UVIT opens a window of opportunity to make a better understanding of the star formation properties of NGC 628 with its $28^{\prime}$ field of view, which covers the galaxy much beyond the optical radius, and that too at a remarkable resolution of $1.4^{\prime\prime}$ \citep{GeorgeJelly2018,Mondal2018, Mondal2021}. The age and mass of the star forming regions are estimated using simple stellar population (SSP) models. We also explored the UV features of the headlight cloud in NGC 628 using UVIT data. The data and analysis are explained in Section \ref{sect:Data&analysis}, followed by the theoretical models in Section \ref{sect:models} and the results in Section \ref{sect:results}. Discussion and a summary is presented in Sections \ref{sect:Discussion} and \ref{sect:summary} respectively. A flat Universe cosmology is adopted throughout this paper with $H_{0}$ = 71 $km s^{-1} Mpc^{-1}$ and $\Omega_{M}$ = 0.27 \citep{Komatsu2011}. In the galaxy rest-frame, $1^{\prime\prime}$ corresponds to a distance of 44.7 pc.

\begin{table}
	\centering
	\caption{Basic parameters of NGC 628}
	\label{tab:Table1}
	\begin{tabular}{lll} 
		\hline
		Parameter & Value & Reference  \\
		\hline
		Type & SA(s)C & 1 \\
		RA (hh mm ss) & 01 36 41.747 & 2 \\
		Dec (hh mm ss) & +15 47 01.18& 2\\
		Distance & 9.6 Mpc & 3 \\
		Inclination (i) & $9^{\circ} $ & 4\\
		Position angle (PA) & $25^{\circ}$ & 5\\
		\hline
	\end{tabular}

\footnotesize{$^{1}$\citet{devac1991},$^{2}$\citet{Evans2010},$^{3}$\citet{Kreckel2017},$^{4}$\citet{Blanc2013},$^{5}$\citet{sakhibov2004}}
\end{table}

\section{Data and Analysis}
\label{sect:Data&analysis}

To understand the mechanisms affecting the star formation in NGC 628 and hence to better understand the secular evolution, we use the UVIT data obtained from the ASTROSAT ISSDC archive. UVIT, with a $28^{\prime}$ field of view, observes simultaneously in FUV (130-180 nm), NUV (200-300 nm), and VISible (320-550 nm) bands. FUV and NUV channels are used for science observations, whereas the primary objective of the VIS channel is to aid the drift correction. Compared to its predecessor GALEX, which has a spatial resolution of $\sim$ $5^{\prime\prime}$, UVIT provides a better resolution of $\sim$ $1.2^{\prime\prime}$ and $1.4^{\prime\prime}$ for NUV and FUV filters, respectively. Even though the UVIT plate scale at the detector plane is 3.33"/pixel, an onboard algorithm enabled these pixels to localize the position of a photon to $1/8^{th}$ of the pixel element. We also made use of the B-band image obtained from the (KPNO) 0.9 m telescope at the Kitt Peak National Observatory for this study (observers: van Zee, Dowell).

\begin{figure*}
\begin{center}
\includegraphics[width = 2\columnwidth]{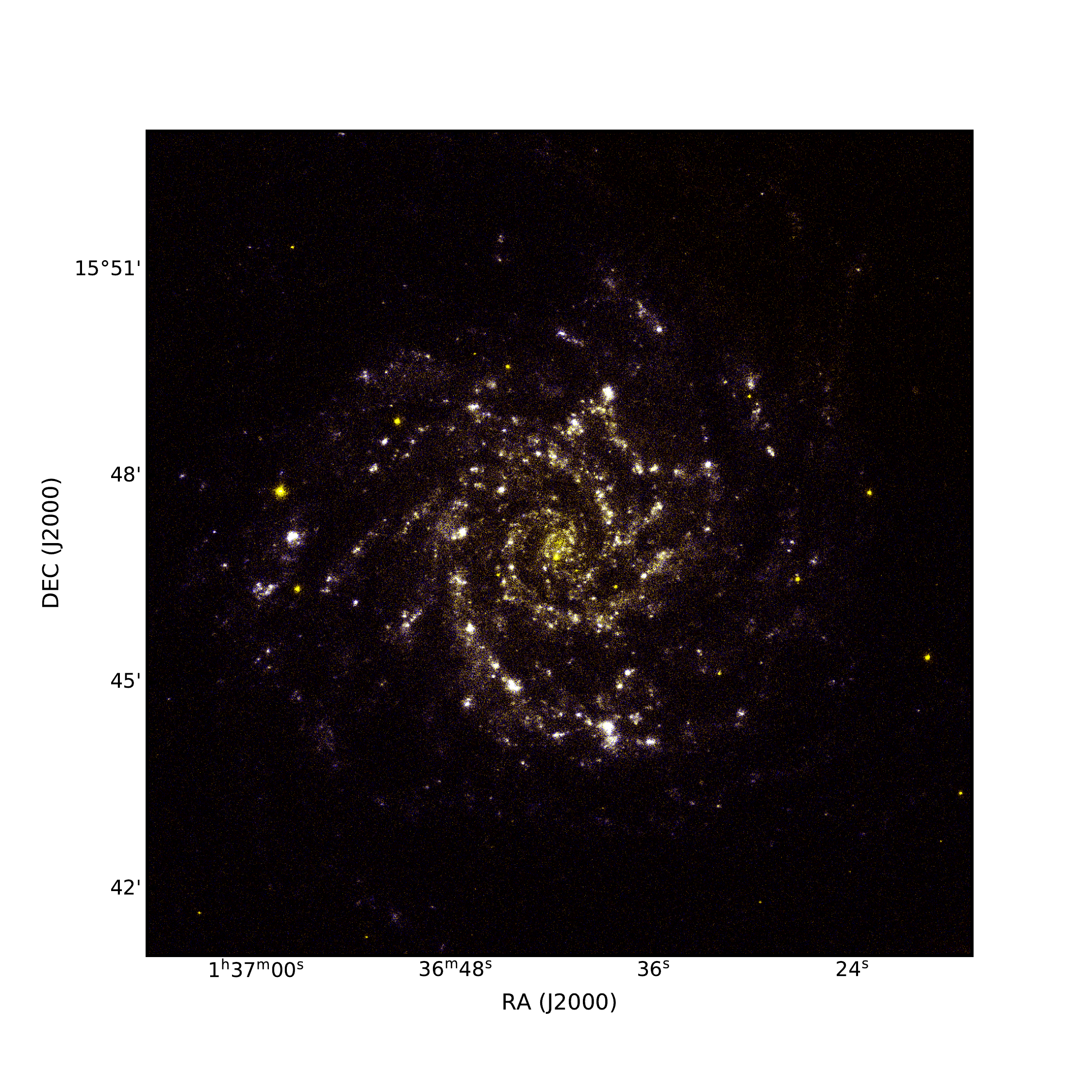}
\caption{UVIT color composite image of the galaxy NGC 628. The emission in F148W and N263M filters are represented by blue and yellow colors respectively. Foreground stars are seen in yellow.}

\label{fig:composite}
\end{center}
\end{figure*}

UVIT has observed NGC 628 (PI: ASK Pati, observation id: G06\_151, date of observation: 29-Nov-2016) in FUV and NUV channels. Each channel of UVIT consists of narrow as well as broadband filters. {In this study, we made use of the FUV F148W filter, with a peak wavelength of 1481 $\AA$ and NUV 263M filter with a peak wavelength of 2632 $\AA$ \citep{Tandon2020} having an exposure time of 1488.7s and 2086.5s respectively}. We employed the software package CCDLab \citep{Postma2017} to reduce the Level 1 data.  Drift correction has been accounted using the NUV images since the drift correction using VIS image resulted in a lesser number of frames than the one performed using NUV images. By making use of the calibration files provided by \citet{Girish2017} and \citet{Postma2011}, each image is flat fielded, followed by the distortion correction and pattern noise. Final images are produced by combining the corrected images using CCDLab itself. The astrometric solutions are also made using the same software. For the FUV and NUV filters the adapted zero point magnitude values are 18.097 and 18.146 respectively \citep[Table 3,][]{Tandon2020}. Figure \ref{fig:composite} represents the UVIT color composite image of NGC 628 generated using the UV emission in F148W and N263M filters.

\section{Theoretical models}
\label{sect:models}

Understanding the properties of star forming regions in a galaxy such as age and mass is the key that enables us to get a clear idea about the process of star formation in the galaxy. In this study, we made use of the Starburst99 SSP model \citep{Leithere1999}
to characterize the young star forming regions of the galaxy NGC 628. Starburst99 is a spectro-photometric SSP model which provides us the spectra of young star clusters for a set of chosen parameters. The output spectra obtained from Starburst99 can be used to analyze the star-forming regions' evolutionary stages.
\citet{Mondal2018, Mondal2021} used these models extensively to estimate the masses of several compact star forming regions identified using UVIT images.

\begin{table}
\caption{Starburst99 model parameters}
 \begin{tabular}{|c|c|}
  \hline
  Parameter & Value \\
  \hline
   Star Formation  & Instantaneous  \\
   Stellar IMF & Kroupa (1.35, 2.35) \\
    Stellar mass limit  & 0.1, 0.5, 120 $M_{\odot}$ \\
    Cluster mass range   & $10^{3}M_{\odot}-10^{7}M_{\odot}$ \\
    Stellar evolution track  & Geneva (High mass loss) \\
    Metallicity & Z= 0.02 \\
    Age range  & 1-900 Myr \\
    \hline
     \end{tabular}
     \label{tab:criteria}
\end{table}

\begin{figure}
\begin{center}
\includegraphics[width = 1\columnwidth]{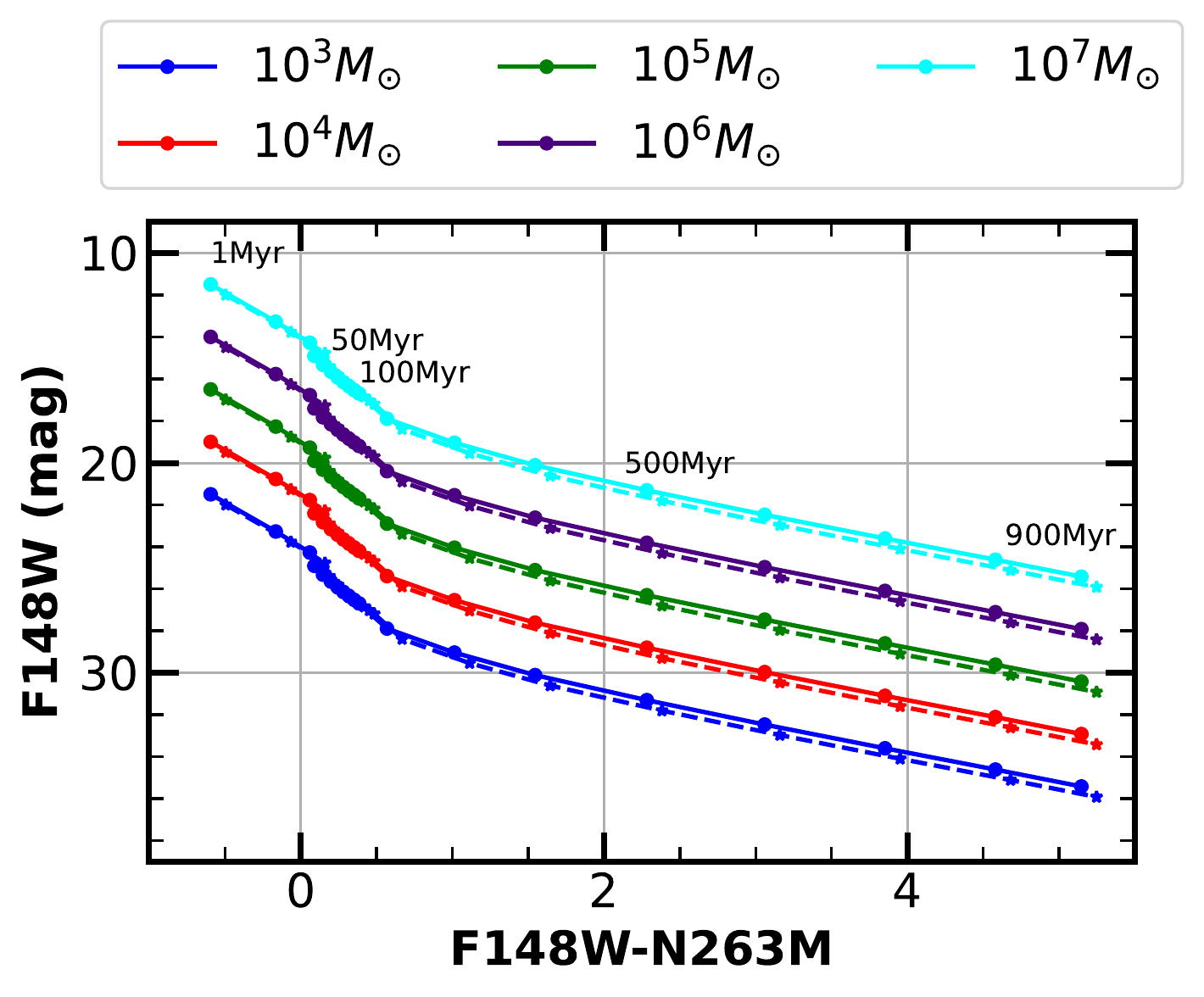}
\caption{Theoretical color-magnitude plot obtained for UVIT filters using Starburst99 models for the cluster mass ranging from $10^{3} - 10^{7}M_{\odot}$ with an age range of 1 to 900 Myr. The dashed lines represents the effect of Milky Way extinction and reddening.}
\label{fig:sb99_mass}
\end{center}
\end{figure}

For this study, an instantaneous star formation law is used along with the Kroupa stellar initial mass function ($\alpha$ = 1.35 and 2.35) within a stellar mass range of 0.1-120 $M_{\odot}$. Assuming solar metallicity, spectra in the age range between 1-900 Myr are generated. The parameters and the input values we used are listed in Table 2. The expected magnitudes of these obtained spectra in F148W and N263M filters have been estimated by convolving with the corresponding UVIT filter effective area curves. This procedure is performed on the spectra obtained for five cluster masses ranging from $10^{3} M_{\odot}$ to $10^{7} M_{\odot}$. To estimate the age and masses of the star forming regions in NGC 628, we used the plot represented in Figure \ref{fig:sb99_mass}, generated using the Starburst99 model. Mass is estimated using the color-magnitude diagram given in Figure \ref{fig:sb99_mass} and age is estimated using the color value.

Extinction plays a significant role when studying the UV images of galaxies since the UV region exhibits a higher value for the extinction coefficient. In order to account for the Galactic extinction, we used the \citet{Cardelli1989} extinction law with $R_{v} = A_{v}/E(B-V) = 3.1$. For the UVIT bands, the ratio of $A(\lambda)/E(B-V)$ is 8.1 and 6.5 for F148W and N263M, respectively. In order to obtain the foreground extinction for UVIT filters from the Milky Way galaxy in the direction of NGC 628, we used the $A_{v}$ value of 0.192 obtained from \citet{Schlafly2011}. From the analysis, $A_{F148W}$ and $A_{263N}$ are found to be 0.5 mag and 0.4 mag, respectively. These extinction values are used to correct the UV magnitudes in both passbands.


\section{Results}
\label{sect:results}
\subsection{Star forming regions in NGC 628}
\label{subsect:regions}
The star forming regions in the galaxy enables us to achieve a better understanding of the evolution of the galaxy. The properties of the regions, such as the size, star formation rate (SFR), radial distance, and distribution along the morphological features, help us to break down the mechanisms driving the evolution of the galaxy. 

\begin{figure}
\begin{center}
\includegraphics[width = 1\columnwidth]{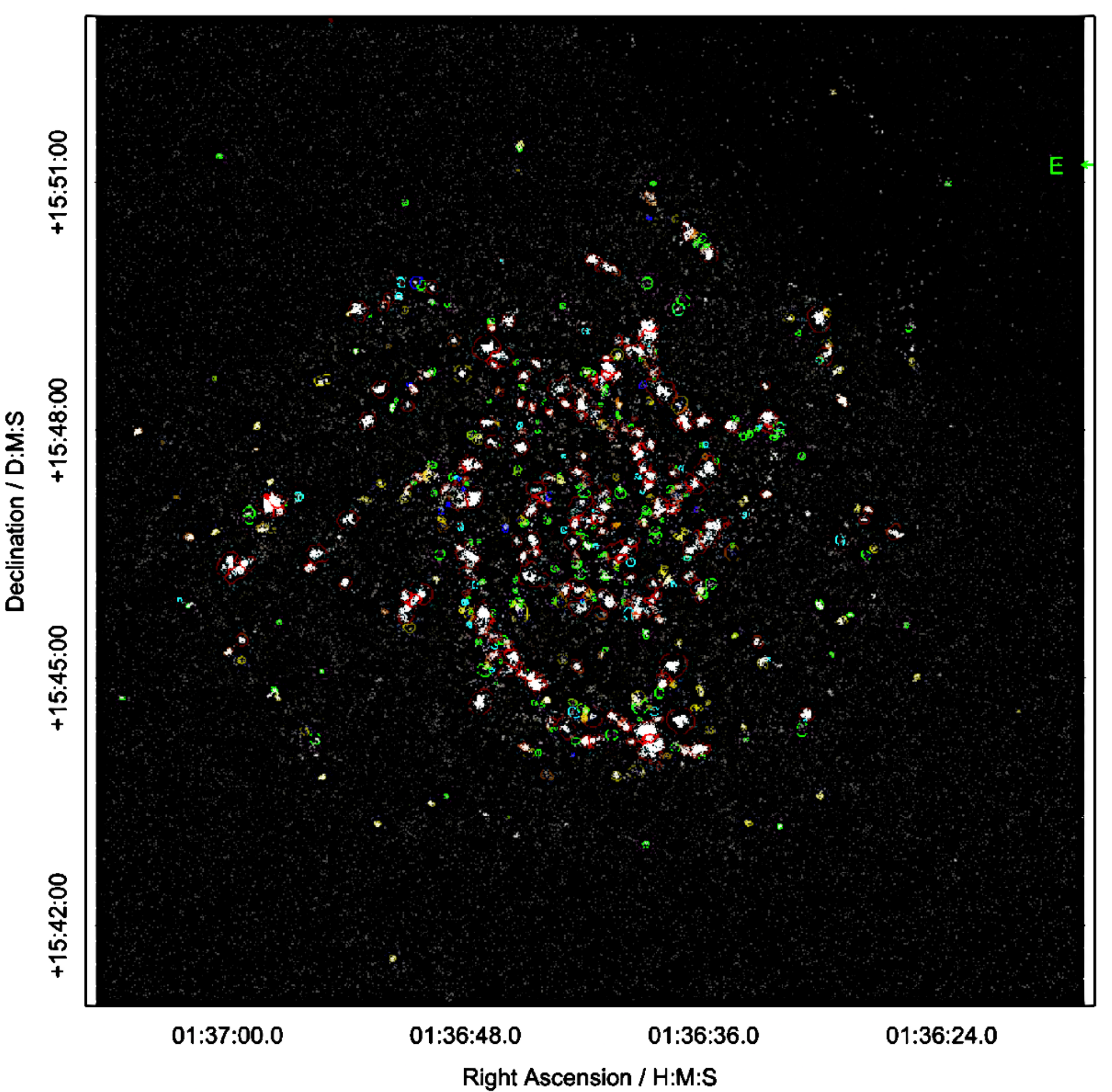}
\caption{Star forming regions identified by ProFound in the UVIT F148W image of NGC 628. Contours of the regions are color coded by ProFound based on the flux contribution from the identified regions.}
\label{fig:profound}
\end{center}
\end{figure}

Since the massive young OBA stars emit predominantly in FUV, it helps us to trace the young massive star forming regions in a galaxy. We made use of the ProFound package to find out the brightest regions from the UVIT FUV images. ProFound is an astronomical data processing tool available in the R programming language. ProFound identifies the peak flux locations in the image and identifies the source segments by means of watershed deblending. The detected segments are then iteratively grown (dilated) to estimate the entire photometry \citep{Robotham2018}. We used the subroutine of the same name in ProFound to identify the star forming regions in the F148W image. We implemented a criterion that the identified star forming region should span over a minimum of 6 pixels. This criterion was chosen by accounting for the resolution of FUV filters. The consideration used here is that we identify the minimum number of pixels to cover the diameter of a circle with size as that of the resolution of the FUV filter. A {\it skycut} of 3 was applied in the ProFound for the identification of star forming regions. Figure \ref{fig:profound} represents the output obtained from ProFound representing the star forming regions.

In Figure \ref{fig:profound}, the contours of the regions are color-coded according to the flux contained in each region. ProFound identified 469 bright regions in the FUV image of NGC 628 and the basic information of the identified regions such as position, magnitude and extent are obtained. After dilating to obtain a total flux measurement, the output of the analysis performed with ProFound gives us the number of pixels contained in the identified segments. We made use of the total number of pixels to estimate the area of the identified star forming regions. The magnitude of each region has been obtained on each of the identified regions. Also, the obtained magnitudes are corrected for line of sight Milky Way extinction. Star forming regions having FUV mag $<$ 21 are only selected for further analysis to exclude the regions with larger photometric error \citep{Mondal2021}. We obtained a final sample of 300 star forming regions in NGC 628.

The internal extinction due to the interstellar medium of galaxies affects the derived parameters such as star formation density (SFRD), age, and mass of the star forming regions. \citet{Sanchez2011} based on the PPAK Wide-field Integral Field Spectroscopy studied the stellar populations and line emission in NGC 628. They derived the dust extinction using the $H_{\alpha}/H_{\beta}$ line ratio. No specific extinction trend has been reported along the spiral arms or in any radial distribution. Since our study is based on the UV regime, which is strongly affected by the extinction, the variable internal extinction needs to be accounted for. The field of view difference between the UVIT images and the  IFS images used in \citet{Sanchez2011}, limits our homogeneous estimation of internal extinction in NGC 628. Hence we use the   Spitzer MIPS $24\mu$m image of NGC 628 obtained from the SINGS data archive \citep{Kennicutt2003} to account for the internal extinction. It needs to be noted that the resolution of UVIT images is ~4 times better than the MIPS $24 \mu$m images. Despite the fact that there is resolution mismatch, MIPS $24 \mu$m images are the best available data set to account for the internal extinction in our study. The segmentation maps obtained for the FUV image are overlaid on the MIPS $24 \mu$m image, and the internal extinction corrected magnitudes are estimated using the relation obtained from \citet[][Table 2]{Kennicutt2012} and is given below:

\begin{equation}
L(FUV_{corr}) = L(FUV_{obs}) + 3.89 \times L(25 ~ \mu m)
\end{equation}

\begin{equation}
L(NUV_{corr}) = L(NUV_{obs}) + 2.26 \times L(25 ~ \mu m)
\end{equation}

Hereafter, we use the extinction corrected magnitudes for further analysis. The star formation rate in each region is estimated using the relation obtained from \citet{Karachentsev2013} and is given below.

\begin{equation}
log(SFR_{FUV}(M_{\odot}yr^{-1})) = 2.78-0.4mag_{FUV}+2log(D)
\end{equation}

\begin{figure}
\begin{center}
\includegraphics[width = 1\columnwidth]{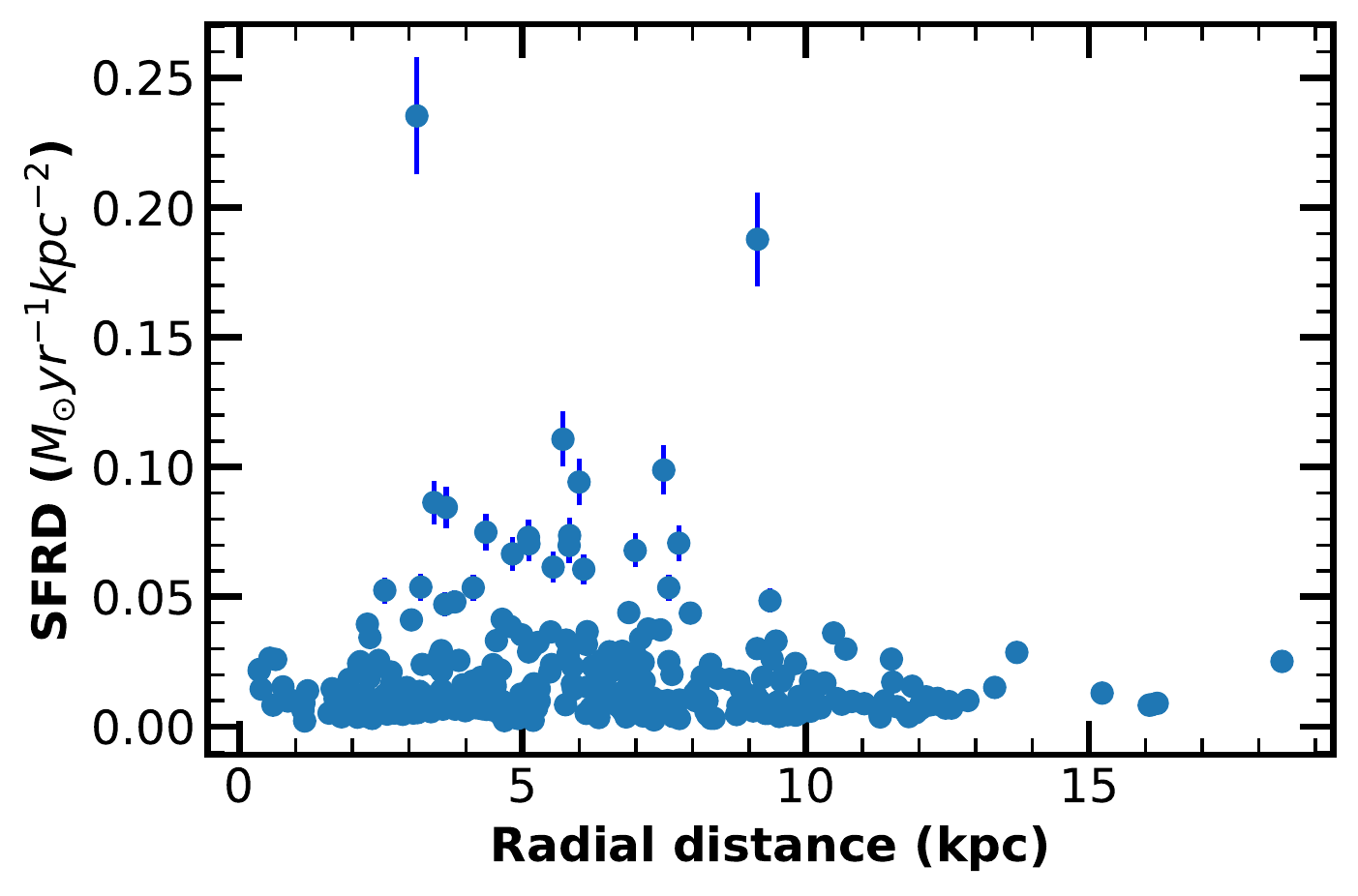}
\caption{Distribution of SFRD of the regions with respect to the radial distance from the center. From the figure it is evident that most of the star forming regions with higher SFRD is situated in the outer regions of the disk.}
\label{fig:SFRD}
\end{center}
\end{figure}

where, $mag_{FUV}$ denotes the background and extinction corrected magnitude and D is the distance to the galaxy in Mpc. 

Figure \ref{fig:SFRD} represents the distribution of SFRD of the star forming regions with respect to the radial distance from the center of the galaxy. From  Figure \ref{fig:SFRD}, ~ it is evident that the regions with SFRD $>$ 0.05 $M_{\odot} yr^{-1} kpc^{-2}$ are situated at a galactocentric distance of 3-10 kpc (outer part of the galaxy). The headlight cloud, which is reported to be an extremely bright HII region in the context of the NGC 628 galactic environment \citep{Herrera2020}, exhibits the highest SFRD of $0.23 M_{\odot} yr^{-1} kpc^{-2}$, in NGC 628. A brief discussion about the headlight cloud is given in Section \ref{sec:Headlight}. To understand the effects of different galactic properties on determining the SFR and the propagation of star formation along the galaxy, we have classified the regions according to their position in the galaxy. The propagation of the star formation along the spiral arms is discussed in Sections \ref{subsect:Arms} and \ref{subsect:Azimuth}.

\subsection{Age distribution}
\label{subsect:age dist}

\begin{figure}
\begin{center}
\includegraphics[width = 1\columnwidth]{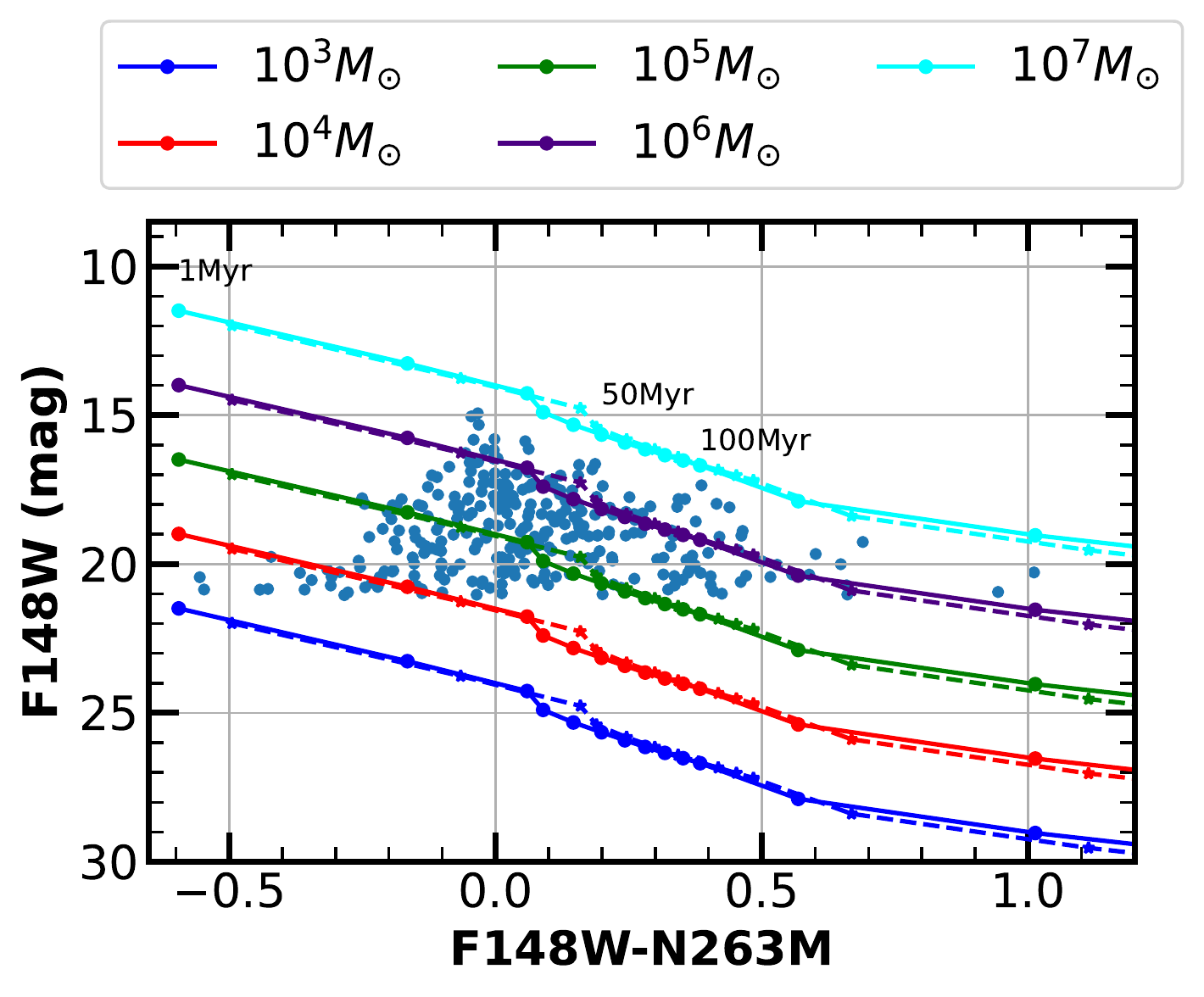}
\caption{UV color-magnitude diagram of the star forming regions (blue circles), over-plotted with the starburst99 model curves. The dashed lines represents the effect of Milky Way extinction and reddening.}

\label{fig:mass_est}
\end{center}
\end{figure}

The star forming regions' age can be estimated from the observed UVIT color using the SSP models generated in Section \ref{sect:models}. The star forming regions identified using ProFound consist of resolved star forming regions, stellar associations, and regions that cannot be further resolved due to the resolution constraints of UVIT. Each of the star forming regions identified by ProFound is considered as a single age stellar population to estimate the age. Using the segmentation maps generated for the FUV images, the corresponding magnitudes in NUV are also obtained from ProFound. From the output, we estimated the extinction corrected magnitudes in F148W and N263M filters and hence the respective color for the regions. Figure \ref{fig:mass_est} represents the UV color-magnitude diagram of the star forming regions, over-plotted with the Starburst99 model curves.
The age of the regions is estimated using linear interpolation along the color axis.

A better understanding of the propagation of star formation across the galaxy can be made from the spatial distribution of age of the star forming regions. Since the evolution of the galaxy can be due to secular and environmental effects, the spatial age distribution provides insights into the factors which affects star formation across the galaxy. The relative distribution of the younger and older star forming regions in the galaxy can be further correlated with the factors affecting the star formation such as the spiral arm and the possible interaction with other galaxies \citep{Gusev2013, shabani2018}.

In order to generate the age map to understand the spatial distribution of age of the star forming regions, we selected seven groups in the age range 1-300 Myr. The selected age groups are 1-20 Myr, 20-40 Myr, 40-70 Myr, 70-100 Myr,100-150 Myr, 150-200 Myr and 200-300 Myr. The bin selection is based on the histogram distribution of the age and the interval is varied with respect to the number of star forming regions in each bin. Also, the selected bin size is larger than the mean error associated with the estimated age in each bin. Left panel of the Figure \ref{fig:age_map} depicts the age map of the star forming regions identified in this study. It suggests that most of the population identified from the FUV image are young. 91$\%$ of the total star forming regions are found to be younger than 100 Myr. ~ 54$\%$ out of the total star forming regions are younger than 20 Myr, which suggests that, recent star formation occurred in the galaxy. From Figure \ref{fig:age_map} it is evident that the outer regions of the galaxy host younger star forming regions compared to the inner part. This could be due to the inside-out growth of the disk \citep{white1991, Brook2006, Munoz2007}.

\subsection{Mass distribution}
\label{subsect:Mass_dist}

The mass of the star forming regions depends on the mass of the parent molecular cloud. Magnitudes and colors of star forming region in F148W and N263M filters are used to estimate the mass of the star forming regions. By making use of the linear interpolation of the F148W magnitude axis of Figure \ref{fig:mass_est} for the observed color of each star forming region, the corresponding mass has been estimated. From Figure \ref{fig:mass_est} it is evident that the mass of the identified star forming regions cover a wide range of values starting from $10^{3}$ to $10^{7} M_{\odot}$ and peaks around $10^{5} M_{\odot}$. A large number of star forming regions falls in the  $10^{5} M_{\odot}$ to $10^{6} M_{\odot}$ mass range. 

\begin{figure*}
\begin{center}
\includegraphics[width =1\columnwidth]{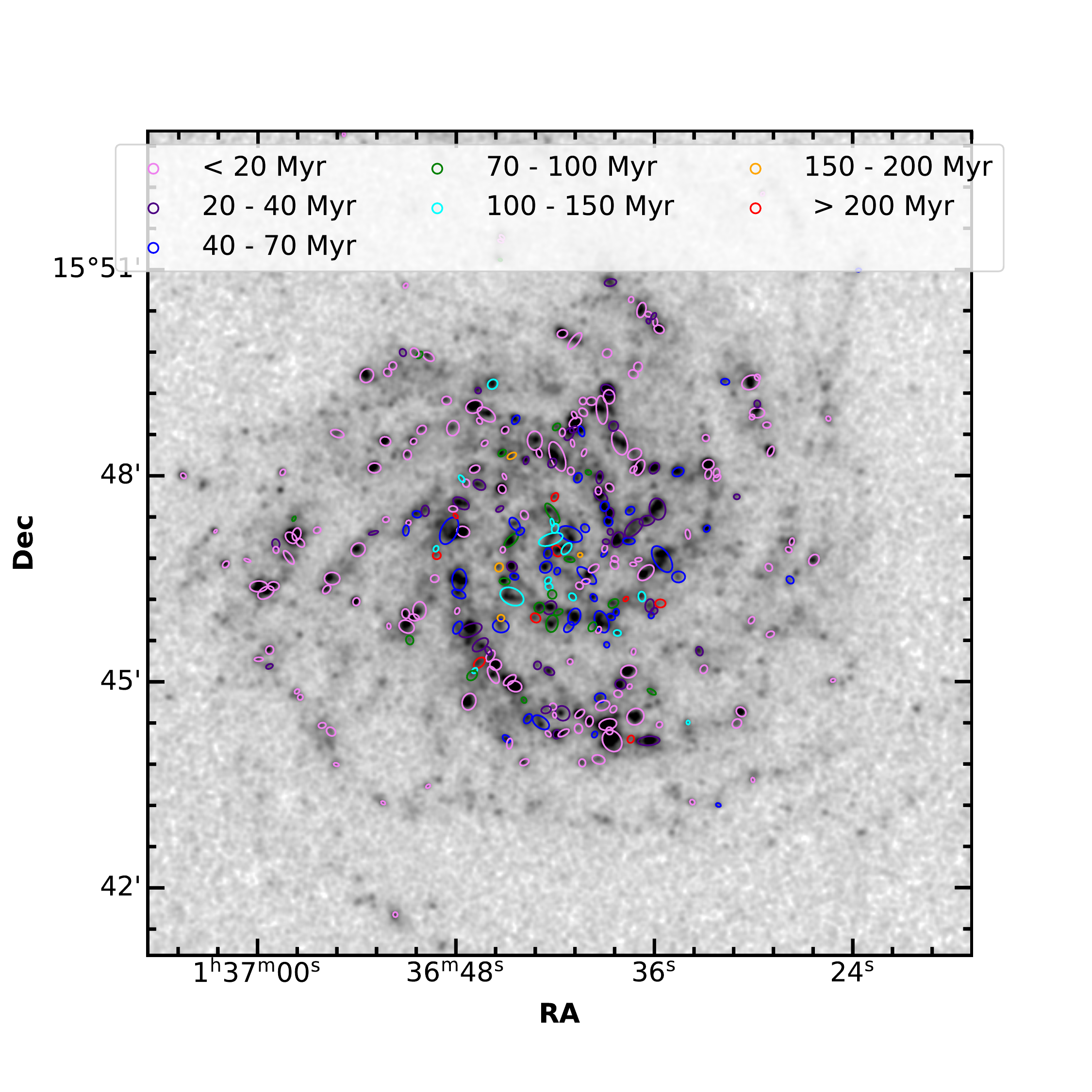}
\includegraphics[width = 1\columnwidth]{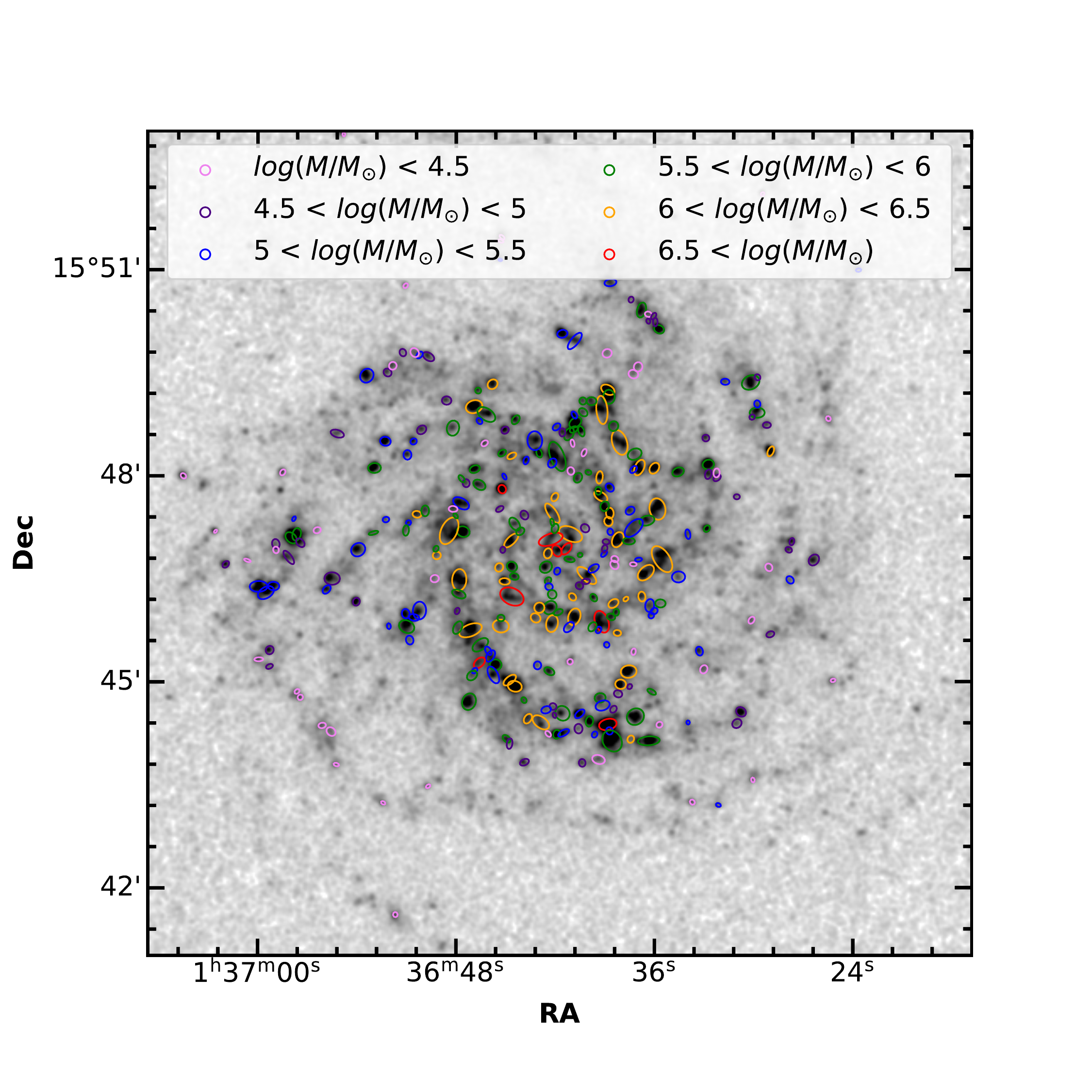}
\caption{Spatial distribution of star forming regions as a function of age (left panel) and mass (right panel) over-plotted on the smoothed FUV image . It is evident that the younger population is located in the outer part of the galaxy whereas the inner part is dominated by older population. Also, most of the massive star forming regions are found to be concentrated towards the center of the galaxy.}
\label{fig:age_map}
\end{center}
\end{figure*}

The distribution of mass of the star forming regions across the galaxy helps us to understand the star formation properties across the galaxy. Right panel of the Figure \ref{fig:age_map} represents the mass map of the star forming regions identified in NGC 628. The mass range selected are log$(M/M_{\odot}) < $ 4.5, 4.5 $ < log(M/M_{\odot}) < $ 5, 5 $ < log(M/M_{\odot}) < $ 5.5, 5.5 $ < log(M/M_{\odot}) < $ 6, 6 $ < log(M/M_{\odot}) < $ 6.5 and $log(M/M_{\odot}) > 6.5 $. The bin selection is performed as discussed in section \ref{subsect:age dist}. Most of the less massive star forming regions (log$(M/M_{\odot}) < $ 4.5) are located in the outer parts of the galaxy. The highly massive star forming regions are situated in the inner part of the galaxy. Figure \ref{fig:age_map} suggests that the recently formed star forming regions are less massive than the older star forming regions in the galaxy.

\subsection{Propagation of star formation along the spiral arms of NGC 628}
\label{subsect:Arms}

The results obtained from the studies by \citet{Gusev2013,Gusev2014} and \citet{shabani2018} suggest two different trends in the star formation properties for the spiral arms of NGC 628. The difference between these two studies is the extent of the spiral arms and the selected star forming regions. The studies initiated by \citet{Gusev2013} considers the shorter arm as one with distortion. Hence only the star forming regions before the distortion starts is used in their analysis. In the case of \citet{shabani2018}, they consider the total extent of both the spiral arms. However, their analysis is incomplete because of the unavailability of data footprints.

\begin{figure}
\begin{center}
\includegraphics[width=1\columnwidth]{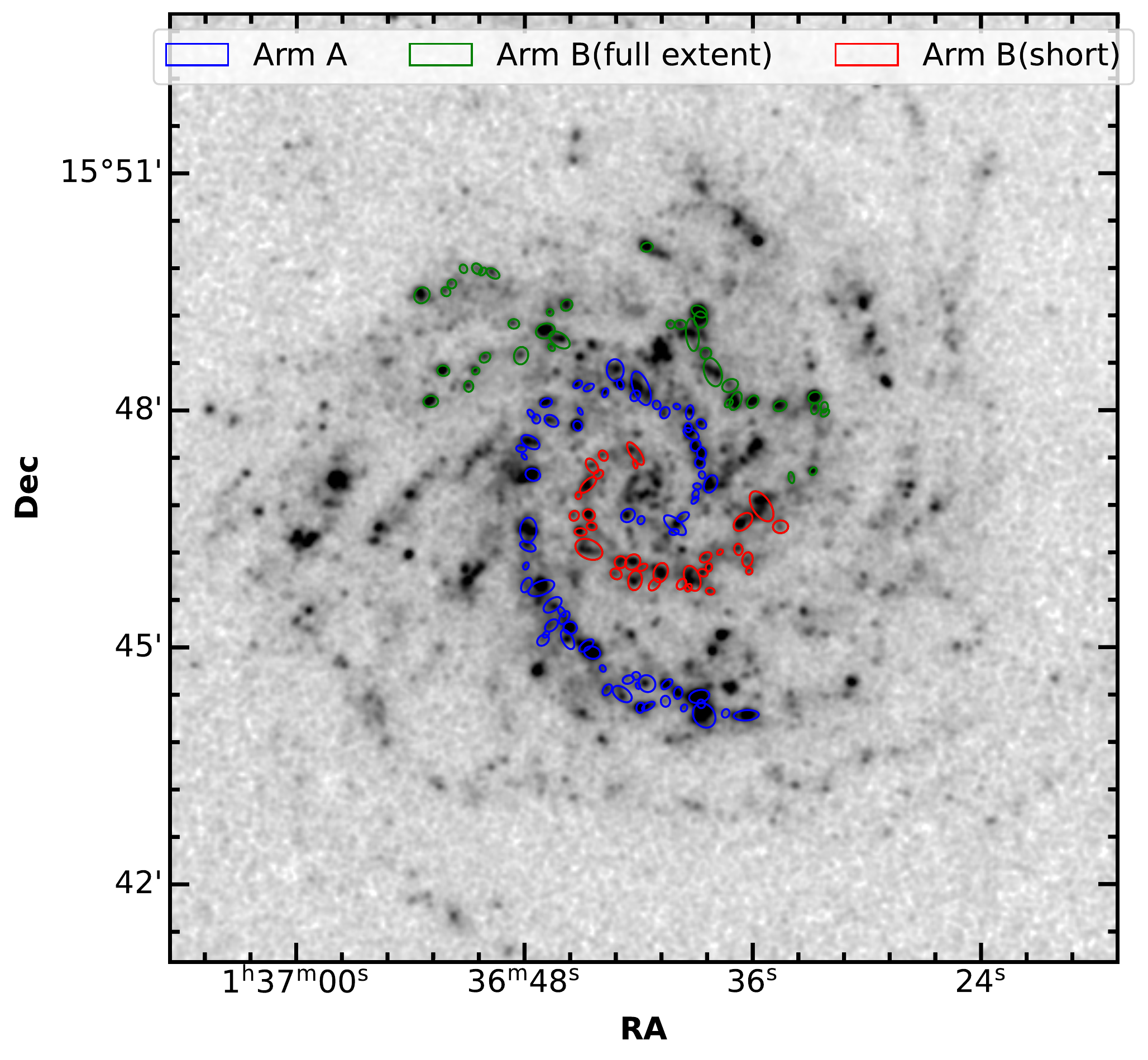}
\caption{Star forming regions selected in spiral arm A and spiral arm B. Blue ellipses represents the star forming regions in the Arm A. Red and Green represents the Arm B, which is considered in two scenarios as discussed in section \ref{subsect:Arms}}
\label{fig:spiral arm}
\end{center}
\end{figure}

In this context, we intend to study the star forming regions detected using UVIT FUV images in the spiral arms of NGC 628. It will help us to understand the difference in the properties, such as SFR, extent, age and mass of the star forming regions. We visually inspected the star forming regions identified using ProFound and separated them to each spiral arm based on their closeness. The arms are mentioned as Arm A (Longer arm/ Arm1) and Arm B (Shorter arm/ Arm2). We consider two scenarios based on the studies by \citet{Gusev2013} and \citet{shabani2018}. In the first case, in Arm B, we only consider the star forming regions before the distorted portion of the arm (as suggested in \citet{Gusev2013} ). In the second case, we consider all the regions identified in Arm B (including the distorted region). In both these cases, the regions selected for Arm A remain the same. Figure \ref{fig:spiral arm} represents the star forming regions identified in the spiral arms.

\begin{figure*}
\begin{center}
\includegraphics[width=2\columnwidth]{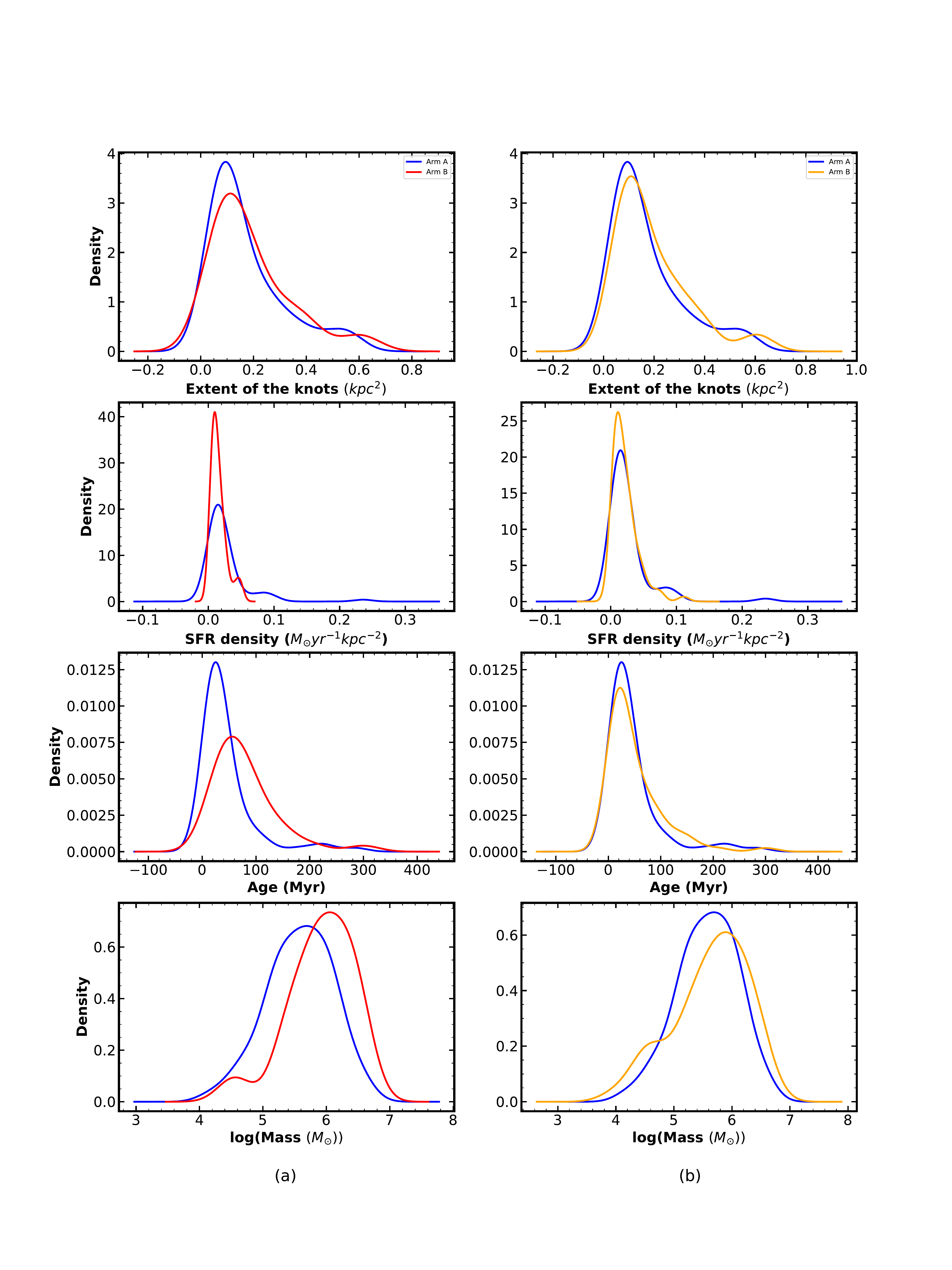}
\caption{Kernel Density Estimate (KDE) plots of properties of the spiral arm A and spiral arm B. In the left panel (a) the regions represented in red in Figure \ref{fig:spiral arm} are only considered for Arm B, whereas the right panel (b) the total extent of Arm B is considered (red $\&$ green ellipses in Figure \ref{fig:spiral arm}).}
\label{fig:combined_age_hist}
\end{center}
\end{figure*}

Figure \ref{fig:combined_age_hist} represents the Kernel density estimate (KDE) plots for the properties of star forming regions such as SFR density, the extent of the regions, age, and mass for both the arms. Left panel (a) represents the first scenario discussed above in which Arm B is considered as the shorter arm. The right panel (b) represents the second scenario in which the full extent of Arm B is considered. From Figure \ref{fig:combined_age_hist}, a significant difference between age and mass of the star forming regions in both the arms is evident in the shorter arm consideration of Arm B.

We performed a two-sample Kolmogorov-Smirnov test (KS test) on the properties of the star forming regions to check whether the properties of Arm A and Arm B constitute a single distribution or not. Same as the KDE analysis, we did the KS test on the samples separately in both the first and second scenarios. KS test performed on the SFRD of the star forming regions suggests that the probability that both the population is part of a single distribution is 34$\%$ if we consider the shorter arm for Arm B. The probability changes to 22$\%$ when we consider the full extent of Arm B. Based on the SFRD estimates, both the samples discard the consideration that there is a significant difference in the star formation rate of the two spiral arms of NGC 628. When the KS test is performed using the age estimates obtained for the star forming regions, the probability value for the shorter arm scenario strongly supports the result of \citet{Gusev2014} that there is a significant difference in the age distribution of star forming regions of longer and shorter arms of NGC 628 (probability = 0.001\%). On the contrary, when we use the Arm B sample with the total extent, the probability value changes to 59$\%$. The higher probability value suggests the population in both arms does not differ much in terms of the age of the regions. While considering the mass of the star forming regions in the arms, the KS test result suggests 0.7\% and 43\% for the shorter arm and total extent considerations of Arm B, suggesting the same result as in the case of age. It could be occurring since the older and massive star forming regions are situated in the inner part of the disk as we discussed in Section \ref{subsect:age dist} and \ref{subsect:Mass_dist}. In both scenarios, the KS test based on the extent of the star forming regions, suggests that both samples only belong to the same population.

\subsection{Azimuthal distance distribution of star forming regions}
\label{subsect:Azimuth}

\citet{Gusev2014} attribute the asymmetric star formation in spiral arms observed in their study to the spiral density waves. \citet{Henry2003} studied the asymmetry in the spiral arms of M51 and suggested that the presence of more than one spiral density wave could be the cause of variable star formation. The presence of a one-armed wave along with the dominant two-armed one in NGC 628 has been proposed by \citet{sakhibov2004}.

\begin{figure}
\begin{center}
\includegraphics[width = 1\columnwidth]{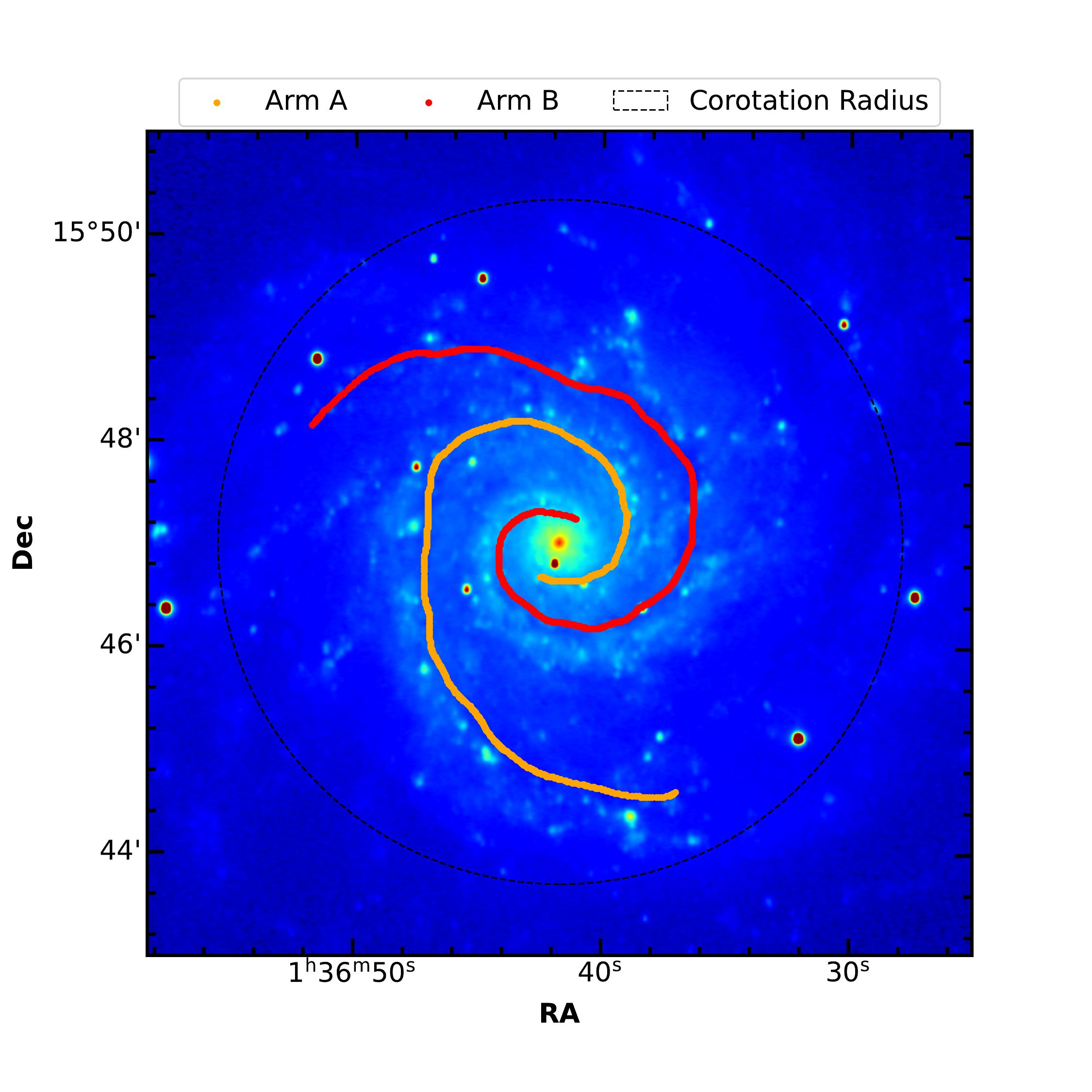}
\caption{Figure shows the optical B band image of the galaxy over-plotted with the ridge lines. The orange and red color lines represents Arm A and Arm B respectively. Dotted line represents the extent of the corotation radius.}
\label{fig:ridge}
\end{center}
\end{figure}

An age gradient across the spiral arms can be used to confirm the the presence of spiral density waves \citep{shabani2018}. To estimate the relative position of the star forming regions with respect to the spiral arms, defining the spiral arms is a requirement. According to the spiral density wave theory, the inner part of the spiral arms forms dark dust lanes due to the compression of gas as it flows through the potential minima of the density wave \citep{Roberts1969}. Since the dust lanes are narrow and well defined in optical images, we use the dust lanes to define the spiral arms of NGC 628. We used the B-band image obtained from the Kitt Peak National Observatory (KPNO) 0.9 m telescope for this purpose (observers: van Zee, Dowell). We defined the spiral arm ridge lines manually in the smoothed B band image of NGC 628. Figure \ref{fig:ridge} represents the B band image over-plotted with the spiral arms selected in both Arms A and B. We adopted a radius of 2 kpc for the bulge-dominated part \citep{shabani2018} and a corotation radius of 7 kpc \citep{sakhibov2004} for NGC 628. To do this analysis we assigned all the star forming regions inside the corotation radius to either Arm A or Arm B based on the location. For each star forming region, the distance to both the ridgelines are estimated. The star forming region is assigned to the arm, from which it has the minimum distance.

\begin{figure}
\begin{center}
\includegraphics[width = 1\columnwidth]{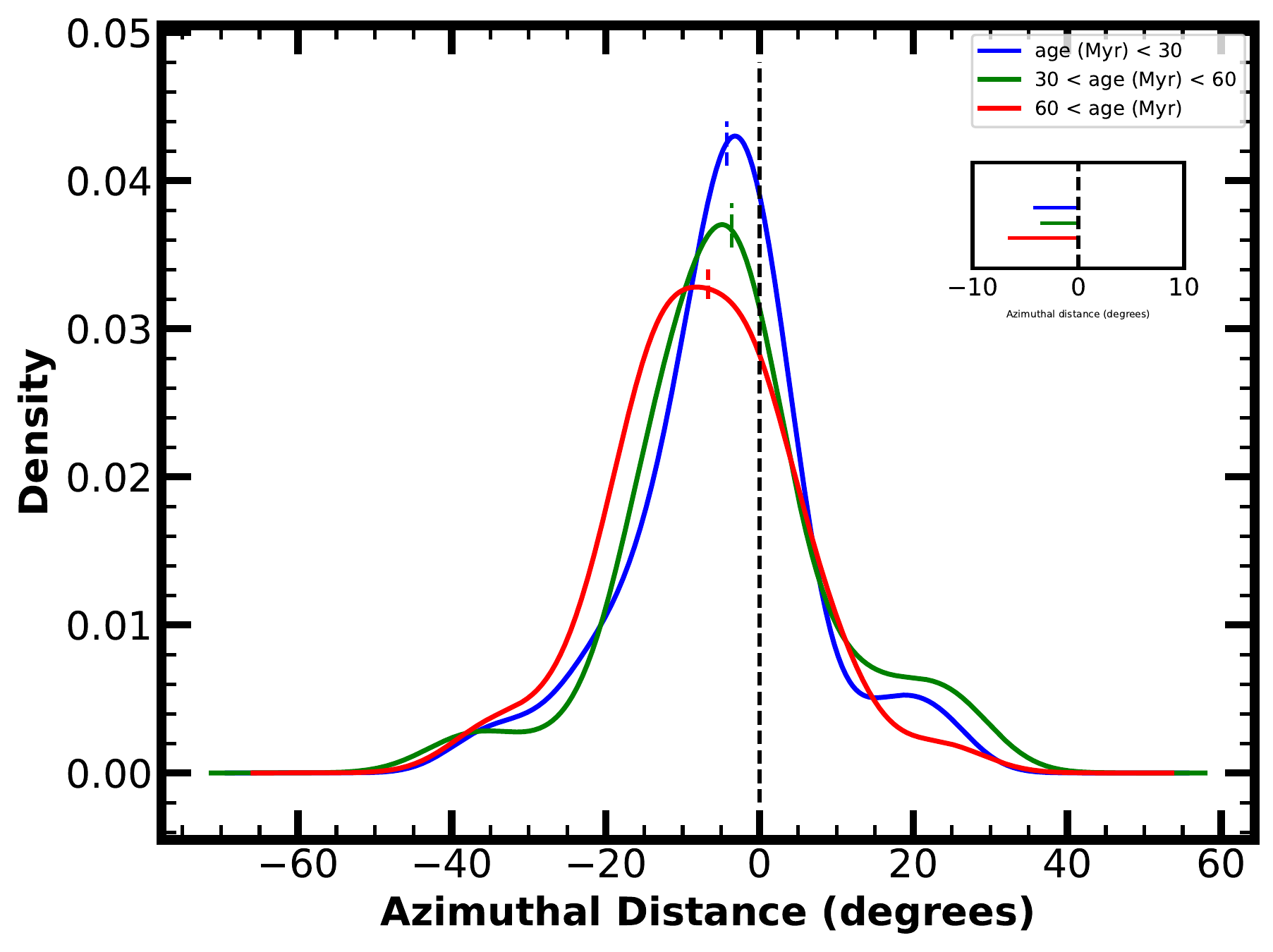}
\caption{KDE plots of azimuthal distance (in degrees) of the star forming regions from their corresponding spiral arm. The median azimuthal distance is shown using the vertical lines drawn in the respective colors. The difference of the median from the ridgeline is represented in the figure given below the label box. No significant gradient is shown by the three populations.}
\label{fig:azimuth}
\end{center}
\end{figure}

 Based on the position of the star forming region and the ridgeline, the azimuthal distance of the star forming region from the ridgeline is estimated. Based on the histogram distribution of the age of the star forming regions, we selected three age bins for this analysis.  The selected age bins are 1-30 Myr, 30-60 Myr and greater than 60 Myr. Figure \ref{fig:azimuth} represents the KDE plots of the azimuthal distance of star forming regions in the above-mentioned age bins. It is to be noted that the azimuthal distance of zero degrees means that the star forming region falls in the ridgeline. The median azimuthal distance from the ridge line is -5.7, -6.3, and -5.5 degrees for young, intermediate, and old populations, respectively. The median values of azimuthal distance are represented in Figure \ref{fig:azimuth} using vertical lines with corresponding colors. The deviation from the ridge line is represented in the inset of Figure \ref{fig:azimuth}. If an azimuthal age gradient is present, we expect an increasing or decreasing trend in the length of the lines representing the deviation from the ridgeline. From Figure \ref{fig:azimuth} it is evident that there is no significant difference in the azimuthal distribution of star forming regions as a function of age. An age gradient is not present across the spiral arms of NGC 628, which in-turn questions the consideration that density waves are the reason for the formation of spiral structures. This result is consistent with those obtained by \citet{shabani2018}, in which they suggest swing amplification as the possible formation scenario for spiral arms of NGC 628 based on the azimuthal distribution of star forming regions with different ages.

\subsection{Head light region in UV}
\label{sec:Headlight} 
NGC 628 hosts a giant molecular cloud in one of its outer spiral arms and is named as the headlight cloud. Studies performed by \citet{Kreckel2016} and \citet{Kreckel2018} using MUSE $H\alpha$ data found out a bright HII region, having luminosity two orders of magnitude brighter than the mean $H\alpha$ luminosity of the HII regions. \citet{Herrera2020} identified the position of this cloud at an offset of (47", 51") and at a radial distance of 3.2 kpc from the center of the galaxy. The cloud exhibits bright infrared emission in Spitzer and Herschel images. The mass distribution function study by \citet{Sun2018} highlighted that those features exhibiting an intense flux are associated with galactic centers or stellar bars. However, the position of this headlight cloud and the absence of any stellar bars makes the headlight cloud a perfect sample to understand the molecular cloud properties in the disk of a galaxy.

\begin{figure}
\begin{center}
\includegraphics[width = 1\columnwidth]{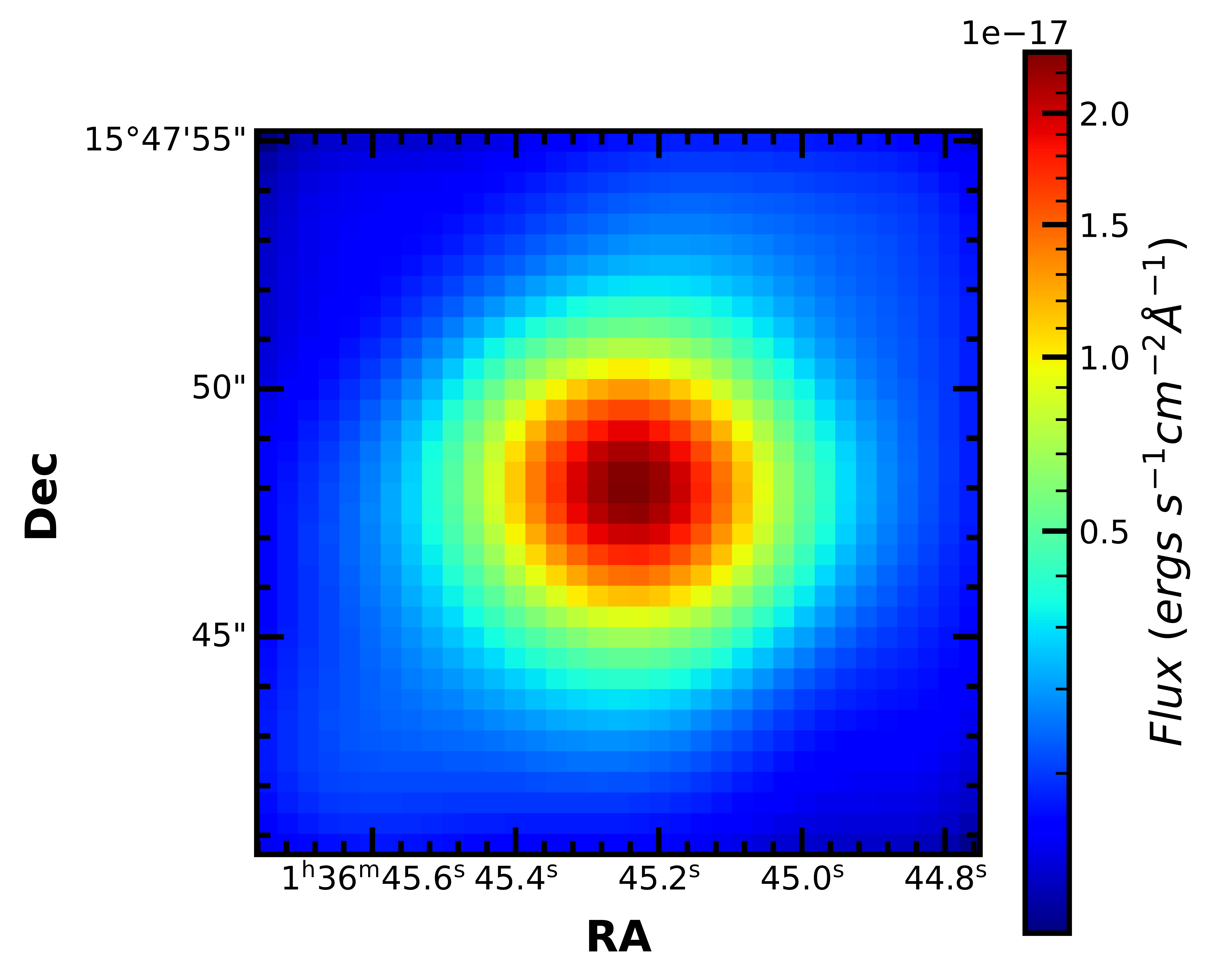}
\caption{FUV image of the headlight cloud of NGC 628 color mapped based on the flux.}
\label{fig:hl}
\end{center}
\end{figure}

Figure \ref{fig:hl} represents the UVIT view of headlight cloud in the NGC 628. \citet{Herrera2020} considers the headlight cloud as the brightest molecular cloud in NGC 628 in their study. In our study also, headlight cloud is found out to be the brightest star forming region in NGC 628. It is to be noted that the 24$\mu$m emission is also high in the headlight cloud region. Further, the age of the headlight cloud is estimated to be 16 Myr. Based on the $H\alpha$ equivalent width (EW) along with the Starburst99 models and using the MUSE spectrum, \citet{Herrera2020} estimated the age of Headlight cloud to be 2-4 Myr. The difference between these two age estimates could be due to the difference in the stellar populations used to estimate the age because of the larger extent of headlight cloud obtained ($\sim$ 280 pc) in this study.

\section{Discussion}
\label{sect:Discussion}

In this study, we primarily focus on identifying the star forming regions in NGC 628 and characterize their properties to make a better understanding of the secular evolution of the galaxy. To attain these goals we used the available UVIT data in F148W and N263M filters. By comparing the flux values obtained in FUV and NUV bands with the theoretical models we estimate the age and mass of the star forming regions. It must be noted that the stochastic sampling of IMF could affect the results obtained from the Starburst99 analysis. Since most of the star forming regions identified in this study have mass greater than $10^{4} M_{\odot}$ we can consider that the IMF is fully sampled, and hence the effect of stochastic sampling of IMF is less likely to affect the results of our study  \citep{daSilva2012}.

Based on the results obtained using FUV and NUV images from UIT, \citet{Cornett1994} found that the central region of NGC 628 does not host a significant population of OB stars. Based on the Effelsberg maps, \citet{Mulcahy2017} found that the northern spiral arm of NGC 628 consists of many HII regions, which are also bright in HI \citep{Walter2008} and infrared images \citep{Kennicutt2011}.  \citet{Mulcahy2017}, based on their analysis along with the results from \citet{Marcum2001}, suggests that within the past 500 Myr, the entire disk of NGC 628 has undergone active star formation. They also concluded that inner regions had experienced a more declining star formation than the galaxy's outer regions. These results are consistent with the results obtained from our study (section \ref{subsect:regions} \& \ref{subsect:age dist}). Our UVIT analysis shows that the inner part of the galaxy hosts an older population, whereas a comparatively younger population dominates the outer part of the galaxy. It suggests an inside-out growth of the galaxy.

The reported asymmetry in the star formation properties of the arms has been attributed to the presence of more than one spiral density wave in the spiral arms \citep{Gusev2014}. From the analysis described in section \ref{subsect:Arms}, it is noted that there is no significant difference in the SFRD estimates in the star forming regions corresponding to Arm A and Arm B. In section \ref{subsect:Azimuth} we have analyzed the azimuthal age gradient across the spiral arms and found out that there is no significant age gradient to the azimuthal distribution of star forming regions across the spiral arms. This suggests that the density wave theory may not be fully responsible for the formation of spiral arms in NGC 628.

It also needs to be noted that five out of nine supernovae remnants (SNR) in NGC 628 \citep[mentioned in][]{Sonba2010} are located in the spiral Arm A. Three of the recent supernovae, SN 2003gd, 2013ej, and 2019krl, are also located in spiral Arm A. On the other hand, arm B does not host any supernovae or SNRs. \citet{Michalowski2020} exclusively studied the SNs 2002ap, 2003gd, 2013ej, and 2019krl located in NGC 628. They found that SN 2002ap is located at the end of an off-centre asymmetric 55 kpc-long HI extension containing 7.5\% of the total atomic gas of NGC 628. Based on this result, they suggested that the birth of the progenitor of SN 2002ap can be attributed to the accretion of atomic gas from the intergalactic medium. They also suggested the possibility of tidally disrupted companions of NGC 628 as the reason for the HI extension. They were unable to explain the possible formation scenario for the other 3 SNs located along the spiral arm A of NGC 628.

The results obtained from this study suggest that the density wave scenario might not be fully responsible for the formation mechanism for the spiral arms of NGC 628. As \citet{shabani2018} suggested, swing amplification can be a possible formation scenario for the spiral arms of NGC 628. A combined effect of density wave and swing amplification is also valid. A detailed study regarding the star formation properties of NGC 628 using multiwavelength data could provide a better picture regarding the same.

\section{Summary}
\label{sect:summary}

A summary of the main results obtained from our study is given below.

\begin{itemize}
    \item In this study, we used the UVIT FUV and NUV observations of NGC 628 to identify and characterize the star forming regions in the galaxy.
    
    \item We identified 300 star forming regions in the UVIT FUV image of NGC 628 using the ProFound package.
    
    \item Around 91$\%$ of the star forming regions are found to be younger than 100 Myr. Only 54$\%$ of the regions are younger than 20 Myr. 
    
    \item The youngest clumps ($<$ 10 Myr) are majorly found in the outer extent of the galaxy whereas the central region hosts most of the older population of stars.
    
    \item Mass range of the identified star forming regions extends from $10^{3}$ -- $10^{7} M_{\odot}$
    
    \item Our study suggests that there is no difference between the star formation properties of the spiral arms of NGC 628.  It contradicts the findings of \citet{Gusev2013}.
    
    \item The results obtained from this study did not support the spiral density wave theory for the formation of spiral arms in NGC 628. Also, the absence of an age gradient is consistent with the results from \citet{Foyle2011} and \citet{shabani2018}. 
    
\end{itemize}

\section*{Acknowledgements}

We thank the anonymous referee for the valuable comments that improved the scientific content of the paper. We thank Joseph Postma for his consistent help during the process of reducing UVIT L1 images and Aaron Robotham for his support while performing the source identification using ProFound. UK thanks Chayan Mondal, Prajwel Joseph, Akhil Krishna, Sudheesh and Arun Roy for their valuable suggestions throughout the course of the work. UK acknowledges the Department of Science and Technology (DST) for the INSPIRE FELLOWSHIP (IF180855).  SSK, RT, and UK acknowledge the financial support from Indian Space Research Organisation (ISRO) under the AstroSat archival data utilization program (No. DS-2B-13013(2)/6/2019). SS acknowledges support from the Science and Engineering Research Board, India through a Ramanujan Fellowship. This publication uses the data from the UVIT, which is part of the AstroSat mission of the ISRO, archived at the Indian Space Science Data Centre (ISSDC).We gratefully thank all the individuals involved in the various teams for providing their support to the project from the early stages of the design to launch and observations with it in the orbit. We thank the Center for Research, CHRIST (Deemed to be university) for all their support during the course of this work.

\section*{Data availability}

The UVIT data used in this article will be shared on reasonable request to the corresponding author. All the data is already available at \url{https://astrobrowse.issdc.gov.in/astro_archive/archive/Home.jsp}.



\bibliographystyle{mnras}
\bibliography{bibtex} 

\end{document}